\documentclass[preprint]{aastex}
\usepackage{amssymb}

\shorttitle{Microquasar jets}
\shortauthors{Hao \& Zhang}

\usepackage{natbib}

\begin{document}
\title{LARGE SCALE CAVITIES SURROUNDING MICROQUASARS INFERRED FROM EVOLUTION OF THEIR RELATIVISTIC JETS}

\author{J. F. Hao\altaffilmark{1} and S. N. Zhang\altaffilmark{1,2,3}}
\affil{jingfang.hao@hotmail.com, zhangsn@mail.tsinghua.edu.cn}
\affil{$^1$Department of Physics and Tsinghua Center for
Astrophysics, Tsinghua University, Beijing 100084, China}
\affil{$^2$Key Laboratory of Particle Astrophysics, Institute of
High Energy Physics, Chinese Academy of Sciences, Beijing 100049,
China} \affil{$^3$Physics Department, University of Alabama in
Huntsville, Huntsville, AL 35899, USA}

\begin{abstract}
The black hole X-ray transient XTE J1550-564 has undergone a strong
outburst in 1998 and two relativistic X-ray jets have been detected
years later with the $\it Chandra$ X-ray observatory; the eastern
jet was found previously to have decelerated after its first
detection. Here we report a full analysis of the evolution of the
western jet; significant deceleration is also detected in the
western side. Our analysis indicates that there is a cavity outside
the central source and the jets first traveled with constant
velocity and then were slowed down by the interactions between the
jets and the interstellar medium (ISM). The best fitted radius of
the cavity is $\sim$0.31 pc on the eastern side and $\sim$0.44 pc on
the western side, and the densities also show asymmetry, of
$\sim$0.034 cm$^{-3}$ on the east to $\sim$0.12 cm$^{-3}$ on the
west. The best fitted magnetic fields on both sides are $\sim$0.5
mG. Similar analysis is also applied to another microquasar system,
H 1743-322, and a large scale low density region is also found.
Based on these results and the comparison with other microquasar
systems, we suggest a generic scenario for microquasar jets,
classifying the observed jets into three main categories, with
different jet morphologies (and sizes) corresponding to different
scales of vacuous environments surrounding them. We also suggest
that either continuous jets or accretion disk winds, or both may be
responsible for creating these cavities. Therefore X-ray jets from
microquasars provide us with a promising method of probing the
environment of accreting black holes.

\end{abstract}

\keywords{jets and outflows - radiation mechanisms: nonthermal-
accretion disks - black hole physics -stars: individual (XTE
J1550-564, H 1743-322, GRS 1915+105, GX 339-4) -X-rays: stars}

\section{INTRODUCTION}
Microquasars are well known miniatures of quasars, with a central
black hole (BH), an accretion disk and two relativistic jets very
similar to those found in the centers of active galaxies, only on
much smaller scales (Mirabel $\&$ Rodr\'{\i}guez 1999). The typical
timescales in these systems are also $10^{5}-10^{7}$ times shorter
than those in quasars, thus evolutions of microquasar jets can be
studied in details. Therefore, microquasar systems have been
considered ideal laboratories for understanding accretion process in
black hole systems and might provide us a good alternative to study
AGN phenomena instead of observing them directly (Massi \& Kaufman
Bernad\'o 2008).

Since discovered in 1992, radio jets have been observed in a series
of BH binary systems and several of them showed apparent
superluminal features. In the two well known microquasars, GRS
1915+105 (Mirabel $\&$ Rodr\'{\i}guez 1999) and GRO J1655-40 (Tingay
et al. 1995; Hjellming $\&$ Rupen 1995), relativistic jets with
actual velocities greater than 0.9$c$ were observed. In some other
systems, small-size ``compact jets", e.g. Cyg X-1 (Stirling et al.
2001), and large scale diffuse emission, e.g. SS433 (Dubner et al.
1998), were also detected.

Among these microquasars, XTE J1550-564 is especially interesting
since it was the first Galactic accretion systems that a fast moving
X-ray jet was detected. The large scale and long existing time of
the jets have also made this source unique and valuable in jet
studies. XTE J1550-564 was discovered with RXTE in 1998 during its
strong X-ray outburst on September 7 (Smith 1998). It is believed to
be an X-ray binary system at a distance of $\sim$5.3 kpc, containing
a black hole of 10.5$\pm$1.0 solar masses and a low mass companion
star (Orosz et al. 2002). Soon after the discovery of the source, a
jet ejection with an apparent velocity greater than 2$c$ was
reported by VLBI team (Hannikainen et al. 2001). In the period
between 1998 and 2002, the source also exhibited strong X-ray
activities but no similar radio and X-ray flares were detected again
during these activities (Tomsick et al. 2003).

With the $\it Chandra$ satellite, Corbel et al. (2002) found two
large scale X-ray jets lying to the east and west of the central
source, which were also in good alignment with the central source.
The eastern jet has been detected first in 2000 at a projected
distance of $\sim$21 arcsec and a position angle E of N of
93.8$\textordmasculine\pm$0.9$\textordmasculine$ from the central
black hole. Its apparent proper motion velocity has dropped from an
average of $\sim$32.9 mas/day between 1998 and 2000 to $\sim$21.2
mas/day during its 2000 observations; its X-ray flux also decayed
quite rapidly. Two years later, it could only be seen marginally in
the X-ray image, while a western counterpart became visible at
$\sim$22 arcsec on the other side. The corresponding radio maps are
consistent with the X-ray observations (Corbel et al. 2002).

From the year 2002 to 2003, there were altogether five observations
of this source in the $\it Chandra$ archive. The eastern jet has
disappeared from the X-ray images in late 2002, while the western
jet could be seen clearly in all the five observations, although the
flux decayed quickly as well. Among these observations, only the
first two have been reported and discussed previously (Kaaret et al.
2003; Wang et al. 2003). We therefore use these additional data
points to better constrain the kinematic and spectral analysis for
the eastern jet, and to extend the analysis to the western jet.

In this paper, we first describe the observational data obtained by
$\it Chandra$ X-ray satellite in section 2 and 3. Then we propose a
model for the system and discuss the kinematic and emission process
in section 4. Finally in section 5 and 6, we apply this model to
another source H 1743-322 and make some further discussions based on
these analysis.

 \placetable{table1}

\begin{deluxetable}{rrrrrrrr}
\tabletypesize{\footnotesize} \tablecolumns{8} \tablewidth{0pc}
\tablecaption{XTE J1550-564 $\it Chandra$
Observations\label{table1}} \tablehead{ \colhead{} & \colhead{} &
\colhead{} & \colhead{} & \multicolumn{2}{c}{Eastern Jet} &
\multicolumn{2}{c}{Western Jet} \\
\cline{5-8}\\
\colhead{Obs} & \colhead{Obs}   & \colhead{Obs} & \colhead{Exposure}
    & \colhead{RA} & \colhead{Dec} &
\colhead{RA} & \colhead{Dec} \\
\colhead{Num} & \colhead{ID}   & \colhead{date} & \colhead{(s)} &
\colhead{(15:51:)}  & \colhead{ (-56:28:)}    & \colhead{(15:50:)} &
\colhead{(-56:28:)}}

\startdata
1 &679  & 2000 Jun 9  &3816  &01.\tablenotemark{a}  &36.7  &  &  \\
2 &1845 & 2000 Aug 21 &5160  &01.4$\pm$2$\times$10$^{-2}$ & 36.7$\pm$1$\times$10$^{-2}$   & & \\
3 &1846 & 2000 Sep 11 &4630  &01.5$\pm$2$\times$10$^{-2}$  & 36.6$\pm$1$\times$10$^{-2}$   & & \\
4 &3448 & 2002 Mar 11 &26118 &02.1$\pm$1$\times$10$^{-2}$  & 37.4$\pm$1$\times$10$^{-2}$    & 56.0$\pm$4$\times$10$^{-3}$  & 33.5$\pm$2$\times$10$^{-3}$\\
5 &3672 & 2002 Jun 19 &18025 &  &      & 55.9$\pm$4$\times$10$^{-3}$ & 33.7$\pm$2$\times$10$^{-3}$ \\
6 &3807 & 2002 Sep 24 &24442 &02.17\tablenotemark{b}  & 37.6    &55.86$\pm$6$\times$10$^{-3}$ & 33.9$\pm$3$\times$10$^{-3}$\\
7 &4368 & 2003 Jan 28 &23680 & &        & 55.82$\pm$6$\times$10$^{-3}$  &33.6$\pm$4$\times$10$^{-3}$ \\
8 &5190 & 2003 Oct 23 &47831 & & &55.74$\pm$6$\times$10$^{-3}$
&33.6$\pm$3$\times$10$^{-3}$
 \enddata
\tablenotetext{a}{Positions of the eastern and the western jets are
corrected after calibration of the central source to the position of
the radio observation as RA = 15:50:58.71 and DEC = -56:28:35.7
(Corbel et al. 2001) and the positions are subject to the 0.3 arcsec
positional uncertainty found by Corbel et al. (2001) on their radio
position. The quoted errors are the statistical errors provided by wavdetect. However, $wavdetect$ did not provide error information
for observation 1.} \tablenotetext{b}{The position of the eastern
jet in observation 6 is determined by the center of the strongest
emission region, as marked by a small circle on fig 2.}

\end{deluxetable}

\section{OBSERVATIONS of XTE J1550-564}
There are altogether eight 2-dimensional imaging observations of XTE
J1550-564 in the $\it Chandra$ archive during June 2000 and October
2003 (henceforth observations 1$\sim$8). These observations were all
made by the Advanced CCD Imaging Spectrometer (ACIS). There are also
grating observations providing 1-dimensional imaging in the archive;
but since our interests are in the jet kinematics analysis, these
are not our focus at this stage.

We have down-loaded all data of these eight observations from the
archive and examined the level 2 event files produced by Standard
Data Processing procedures to make our analysis. Exposure time for
these observations ranges from a few kiloseconds (observations
1$\sim$3) to a few tens of kiloseconds (observations 4$\sim$8)
(Table 1). Background light curves are extracted for each of the
observations and only in observation 5 a high background flare is
detected. Except for this period, almost all valid exposure time
intervals of these observations are included in our work.

In the section 2 and 3, we limit most of our analysis to the
kinematic and spectral properties of the western jet. The central
accreting source has already attracted much attention in recent
years and there were already plenty of studies in details in
literature (Corbel et al. 2001; Jain et al. 2001; Kubota $\&$
Makishima 2004; Yuan et al. 2007). The eastern jet has also been
studied fully in 2003 when it was discovered (Tomsick et al. 2003;
Kaaret et al. 2003). As a result, we do not discuss much of our
results for observations 1$\sim$3 here; only comparisons with the
previous works will be noted for consistency. A complete study of
the kinematic and light curve evolution of the western jet is our
main goal in these two sections.

We list the basic information of observations 1$\sim$8 in Table 1,
including the observation ID, date, the positions of the central
point source and the eastern and western jets respectively (in J2000
coordinate). The positions are obtained by the $\it Chandra$
Interactive Analysis of Observations (CIAO) routine $wavdetect$
(Freeman et al. 2002), a commonly used program in determining X-ray
sources in $\it Chandra$ images. From this table, we could see
clearly that an X-ray emission source is detected to the east of the
central source in the first four observations and another source is
detected to the west in the last five observations. Calculations
also show that these two sources, when presented in a single
combined image, are in good alignment with the central compact
object with a position angle E of N of -85.9\textordmasculine
$\pm$0.3\textordmasculine . In observations 5 and 6, no X-ray source
is detected by $wavdetect$ at the position of the eastern jet.
However, from the Gaussian smoothed images (Fig. 2), a weak source
could be recognized in observation 6. We thus select the strongest
emission region of this source manually with a small circle of the
radius of 0.7 arcsec and use the position of its center as one data
point in our further analysis.

In Fig.1, the raw, unsmoothed images of the last five observations
are shown. An extended and moving source can be seen easily in these
images. Its elongation points towards the central source, and has
thus indicated clearly the relation between them. In most cases the
flux from this western source is comparable to the central source;
however in observation 5, the observed surface brightness from the
western source is even higher than the central source (also see
section 3).

\begin{figure}
\includegraphics[width=10cm]{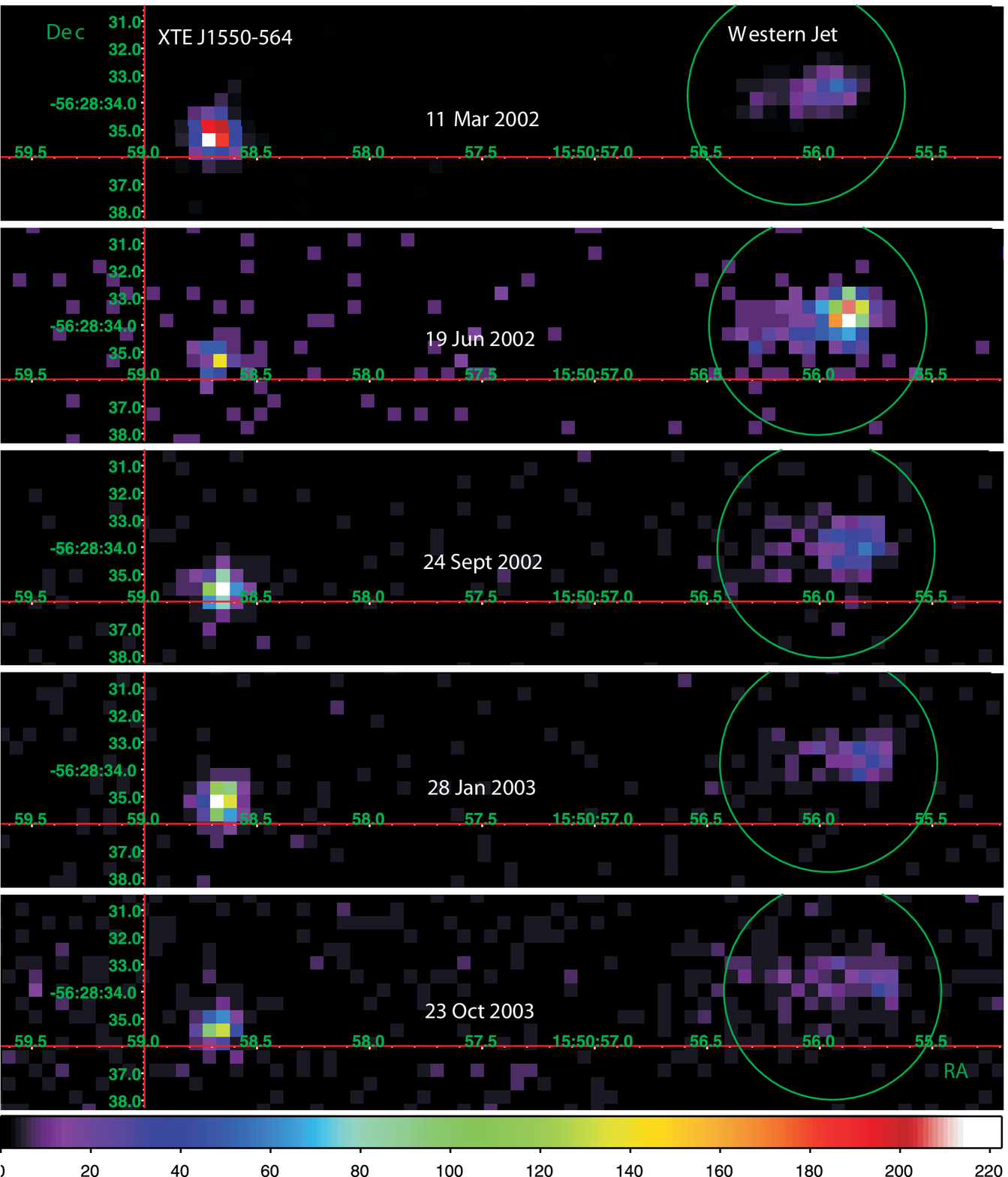}
\caption{The $\it Chandra$ 0.3-8 keV raw image showing XTE J1550-564
and the western jet. The images are in linear scale and no count
saturation has been set. Each image is normalized to its own color
scale. The color bar below the images indicates the relative
strength. The numbers under the color bar indicate the number of
photons in each pixel for the first image (obs 4, ID=3448), and the
peak count values in the other four images (obs 5-8) are 27, 60, 58,
36, respectively. The pixel size is $0".492\times0".492$. The green
circle in each observation shows the spectral extraction region for
each jet, which is of 8 arcsec in diameter and includes almost all
of the jet counts (albeit that there is no guarantee that all jet
photons are included, due to the wing of the point spread function
of the {\it $\it Chandra$} mirror assembly.). The images are aligned
by RA (x-axis).} \label{fig1}
\end{figure}

\placefigure{fig1}

\begin{figure}
\includegraphics[width=10cm]{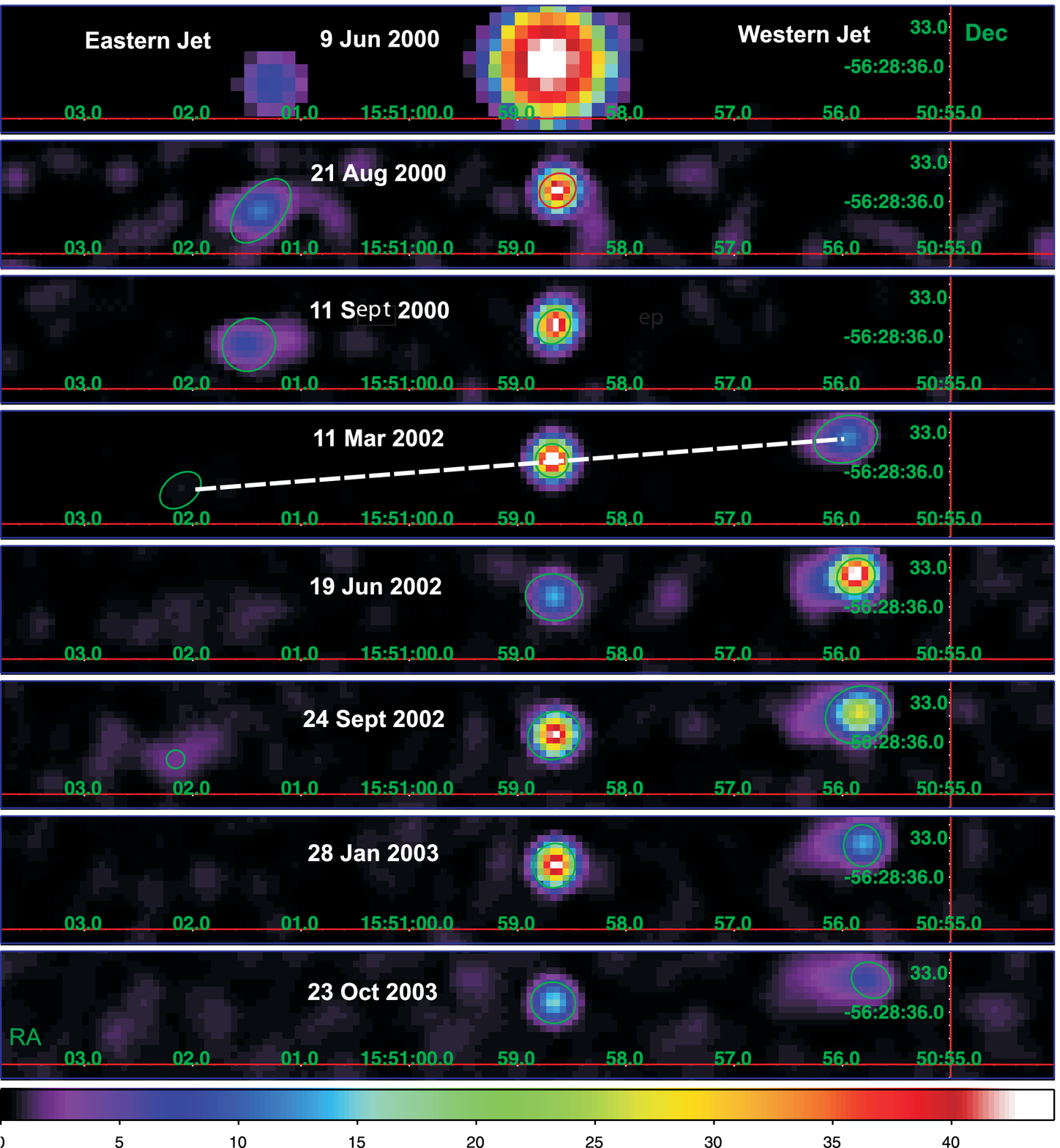}
\caption{The smoothed $\it Chandra$ X-ray images of the eight
observations of XTE J1550-564 and the two jets together. Except for
the first observation, in which the central BH is too bright so a
log scale is set to reveal the eastern jet, all the images are in
linear scale and no count saturation has been set. The appearance
and disappearance of the eastern and western jets can be seen
clearly in the image. Each image is normalized to its own color
scale. The color bar below the images indicates the relative
strength. The numbers under the color bar indicate the number of
photons in each pixel for the fourth image (obs 4, ID=3448) (to be
consistent with fig. 1), and the peak count values in the other
seven images (obs 1-3, 5-8) are 18.8, 3.8, 6.3, 10.3, 13.4, 15.2,
9.0, respectively. The green elliptical regions are source emission
regions detected by {\it wavdetect}. Observation 4 shows the good
alignment of the two jets and the central source. The images are
aligned by RA (x-axis).}
 \label{fig2}
\end{figure}

\placefigure{fig2}

With the data in Table 1, we can calculate the angular separations
between the jets and the central source as well as the average
proper motions. In this paper, we have adopted the assumption that
both the eastern and western jets are related to the same outburst
in September 1998. Except for the outburst occurred in 1998, there
were some other outbursts detected afterwards, including a major
outburst in 2000 and some mini-outbursts in 2001 to 2003 (Tomsick et
al. 2003 and references therein). However, the long term RXTE-ASM
light curve shows that the X-ray flux of the 1998 outburst is four
times higher than that of the 2000 outburst and more than twenty
times higher than that of all the rest (Sturner $\&$ Shrader 2005).
The complex behaviors found in the 1998 outburst, such as irregular
light curves and multiple state transitions, were not detected in
other outbursts. The two X-ray jets are also found to be aligned
with the 1998 radio jets (Corbel et al. 2002), which is also
consistent with the assumption of a common origin of the two jets,
but by itself does not provide definite evidence for the assumption
unless the source is precessing strongly. Based on the above
arguments, we conclude that it is reasonable to assume that both the
eastern and western jets originated from the same 1998 outburst.

Our calculation results are listed in Table 2. A distance of
approximately 23 arcsec is obtained for the western jet with a
slowly increasing value; errors are estimated following the method
described by Tomsick et al. (2003). We calculate the source centroid
for the central source and the X-ray jet respectively and for all
the five observations, the changes of the newly calculated position
with the previous results are less than 0.5\arcsec. Therefore, an
upper limit of 0.5\arcsec is set for the error of the jet distance
(except the eastern jet in observation 6, where a larger error of 1
arcsec was set because of the manual selection method we applied).
We divide the net angular separation by the time interval between
the neighboring observations to estimate the average proper motion
for the jets, and from the results, an approximate estimate of
deceleration could be seen for both jets. The absolute astrometry of
Chandra is calibrated by fixing the X-ray position of the central
source as its radio position reported by Corbel et al (2001).

\placetable{table2}

\begin{deluxetable}{rrrrrrr}
\tabletypesize{\footnotesize} \tablecolumns{7} \tablewidth{0pc}
\tablecaption{Angular Separations and Proper Motions of the Eastern
and Western Jets\label{table2}} \tablehead{ \colhead{} & \colhead{}
& \colhead{Time after} & \multicolumn{2}{c}{Separations to the BH
(arcsec)\tablenotemark{a}} &
\multicolumn{2}{c}{Average Proper Motion (mas/d)\tablenotemark{b}}  \\
\cline{4-5} \cline{6-7}\\
\colhead{Obs Num} & \colhead{Obs ID}   & \colhead{X-ray burst
(days)}    & \colhead{Eastern Jet} & \colhead{Western jet} &
\colhead{Eastern Jet} & \colhead{Western jet} }

 \startdata
1 &679  & 628 &21.5$\pm$0.5   &    &33.9$\pm$0.8  & \\
2 &1845 & 700 &22.8$\pm$0.5   &    &21.1$\pm$0.7  &   \\
3 &1846 & 720 &23.4$\pm$0.5   &    &31.5$\pm$0.7  &   \\
4 &3448 & 1265 &28.6$\pm$0.5   &22.6$\pm$0.5   &9.4$\pm$0.4 &17.9$\pm$0.4\\
5 &3672 & 1365  &     &23.2$\pm$0.5   &    &6.3$\pm$0.4 \\
6 &3807 & 1462  &29.2$\pm$1.4  &23.4$\pm$0.5  & 3.3$\pm$0.3 &1.6$\pm$0.3\\
7 &4368 & 1588  &     &23.7$\pm$0.5   &    &2.2$\pm$0.3 \\
8 &5190 & 1856  &     &24.5$\pm$0.5   &    &3.2$\pm$0.3
 \enddata

\tablenotetext{a}{Errors are estimated by following Tomsick et al.
(2003).} \tablenotetext{b}{Values obtained by dividing the net
angular separation and the error by the time interval between the
neighboring observations.} \tablenotemark{c}{The error bar of the angular separation of the eastern jet in observation 6
is estimated as 2 times the radius of the circle.}

\end{deluxetable}

\section{ENERGY SPECTRUM and FLUX}

We extract the X-ray spectrum in 0.3-8 keV energy band for each
observation. To be consistent with previous analysis of the eastern
jet, we follow some of those previous procedures. Because of the
much longer exposure time, the total counts collected from each of
the last five observations are much more than those from the first
three. We use a circular source region with a radius of 4\arcsec, an
annular background region with an inner radius of 5\arcsec and an
outer radius of 15\arcsec, for each observation. The photon number
inside each selected region is about a few hundreds, possible for
spectral analysis, compared to only $\sim$20 counts in the jet
region in observations 1, 2 and 3. We perform background subtraction
but did not make independent background spectra.

Instrument response matrices (rmf) and weighted auxiliary response
files (warf) are created using CIAO programs {\it mkacisrmf} and
{\it mkwarf} and used in the fitting in {\it Xspec}. We re-bin the
spectra with 10 counts per bin and fit them in {\it Xspec}. Several
spectral models are tested, but the statistical quality is not high
enough to distinguish between them completely. Taking observation 4
for example (for the reason that the photon number of this
observation is the largest), synchrotron model (powerlaw) and self
Compton scattering (compsl) seem to work equally well with the
corresponding reduced $\chi^{2}$ of 0.924 and 0.925. Their
combination (powerlaw+compsl) does not improve the fitting
($\chi^{2}\sim0.981$), but creates a large uncertainty for each
parameter (the best fitting photon index is $2\pm28$ in the combined
model).

In table 3, the results of spectra fitting to the western jet with
an absorbed power-law model are shown (spectra shown in Fig. 3). The absorption column density
is fixed to the Galactic value in the direction of XTE J1550-564
obtained by the radio observations ($N_{H}=9\times10^{21}$cm$^{-2}$)
(Dickey \& Lockman 1990). The first two observations (obs. 4 $\&$ 5)
have been analyzed by Kaaret et al. (2003), and our results are
quite consistent with their work. The mean photon index of the five
observations is around 1.8, also consistent with the previous work
on the eastern jet (Tomsick et al. 2003). A slightly different
photon index is obtained for observation 6 and 8; however, the value
of 1.8 is still within the 1-$\sigma$ error range. The calculated
absorbed energy flux in 0.3-8 keV band is comparable to the value of
the eastern jet. The observed flux decayed from
$\sim1.9\times10^{-13}$ erg cm$^{-2}$ s$^{-1}$ in March 2002 to only
one sixth of this value in October 2003 (see section 4.2). For
observation 6, we also try to fit the spectrum of the eastern jet.
However, the data points are just too few to yield a satisfactory
result. With only 17 counts between 0.3 to 10 keV inside a
8-arcsec-diameter circle, we obtained a reduced $\chi^{2}$ of only
0.13 when using the chi-squared statistic in fitting. When changing
to the Cash statistic, a more suitable method for low counts cases,
the resulting photon index is 1.11 and the estimated flux is
1.6$\times$10$^{-14}$ erg cm$^{-2}$ s$^{-1}$, even higher than the
value in observation 4. We consider this result not very convincing,
or the re-heating process is more complicated than a simple shock.
Complete understanding of this problem is beyond the scope of the
current work. As a result, we drop this spectral data point in the
light curve fitting.

\placetable{c}

\begin{figure}
\includegraphics[height=18cm]{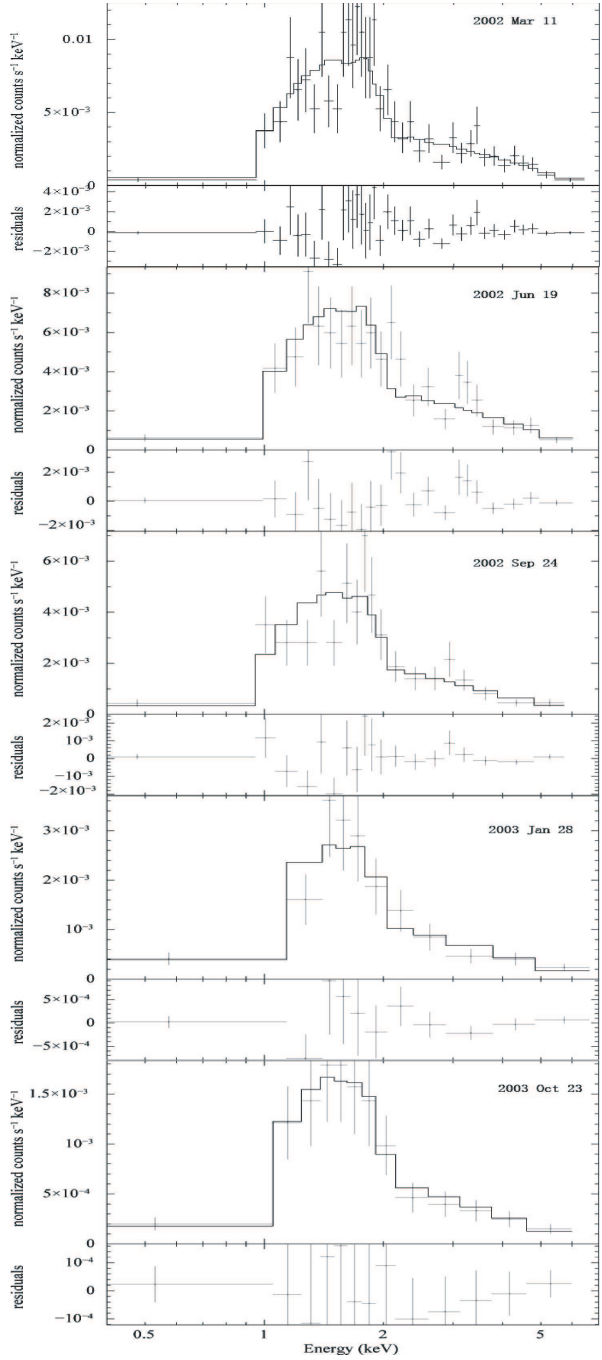}
 \caption{X-ray spectra of emission from the
western jet during March 2002 and October 2003. The five panels are
data from observations 4$\sim$8 respectively. An absorbed power-law
model with the neutral hydrogen column density fixed to the Galactic
value, $9\times10^{21}$cm$^{-2}$, is used in the fitting. Values of
all fitted parameters are listed in Table 3.}
 \label{fig3}
\end{figure}

\placefigure{fig3}

\placetable{table3}

\begin{deluxetable}{rrrrrrr}
\tabletypesize{\footnotesize} \tablecolumns{7} \tablewidth{0pc}
\tablecaption{Spectral Properties of the Western Jet\label{table3}}
\tablehead{ \colhead{} & \colhead{} & \colhead{} & \colhead{} &
\multicolumn{3}{c}{Powerlaw Fitting
Results\tablenotemark{a}} \\
\cline{5-7}\\
\colhead{Obs Num}& \colhead{Time after X-ray}  & \colhead{Total
Counts} & \colhead{Counts Rate} & \colhead{Photon} &
\colhead{Reduced}
& \colhead{Flux\tablenotemark{b}}\\
\colhead{}& \colhead{burst (days)}  & \colhead{(0.3-8 keV)} &
\colhead{(cts/s)} & \colhead{Index}  &\colhead{$\chi^{2}$}&
\colhead{ergs cm$^{-2}$ s$^{-1}$} }

 \startdata
4 &1265   &419 &0.016   &1.75$\pm$0.11  &0.91  &$(1.9\pm0.4)\times10^{-13}$\\
5 &1355   &248 &0.014   &1.71$\pm$0.15  &0.91  &$(1.6\pm0.3)\times10^{-13}$\\
6 &1462   &197 &0.008   &1.94$\pm$0.17  &0.90  &$(8.6\pm1.5)\times10^{-14}$\\
7 &1588   &114 &0.005   &1.81$\pm$0.22  &0.80  &$(5.5\pm1.0)\times10^{-14}$\\
8 &1856   &137 &0.003   &1.97$\pm$0.20  &1.76  &$(3.1\pm0.6)\times10^{-14}$\\
 \enddata

\tablenotetext{a}{Absorbed Powerlaw model with fixed $N_{\rm
H}=9\times10^{21}$ cm$^{-2}$.} \tablenotetext{b}{Absorbed flux in
0.3-8 keV band.}

\end{deluxetable}

\section{JET MODEL}

We now describe the model of Wang et al. (2003), and apply it to the
data, demonstrating how it needs to be revised to fit the full set
of observations of the eastern and western jets. The kinematic and
light curve models are constructed according to the external shock
model in the standard gamma-ray burst (GRB) afterglow theory. In
this model, the kinematic evolution and light curve of the jets can
be interpreted by the interactions between the jet materials and the
interstellar medium (ISM). The relativistic ejecta losses its energy
to the ISM and is therefore slowed down to non-relativistic phase.
In Wang et al. (2003), they used the reverse shock emission model to
fit the light curve of the eastern jet successfully. We therefore
apply this model to the analysis of the western jet as well.

\subsection{Kinematic Model}

As we know, in the external shock model for afterglows of GRBs, the
kinematic and light curve evolution could be understood as the
interaction between the outburst ejecta and the surrounding ISM
(Rees \& M\'{e}sz\'{a}ros 1992). Since microquasar systems are also
Galactic sources embedded inside the ISM, we may infer that the jets
of microquasars should also encounter such interactions and thus be
decelerated during their expansion. The large scale and long
existing time of XTE J1550-564 jets have provided us a good
opportunity to test this scenario.

Because of the good alignment of the two jets and the central
accreting object, we adopt the model of a collimated conical beam
with a half opening angle $\theta_{j}$ expanding into the ambient
medium with the number density $n$. The initial kinematic energy and
Lorentz factor of the outflow material are $E_{0}$ and $\Gamma_{0}$,
respectively. Shocks should arise as the outflow moves on and heat
the ISM, and its kinematic energy will turn into the internal energy
of the medium gradually. Neglect the radiation loss, the energy
conservation equation writes (Huang, Dai, \& Lu 1999):
\begin{equation}\label{a}
(\Gamma-1)M_{0}c^{2}+\sigma(\Gamma_{\rm{\tiny sh}}^{2}-1)m_{\rm{\tiny SW}}c^{2}=E_{0}.
\end{equation}

The first term on the left of the equation represents the kinematic energy of the
ejecta, where $\Gamma$ is the Lorentz factor and $M_{0}$ is the mass of the original
ejecta. The second term represents the internal energy of the swept-up ISM, where
$\Gamma_{\rm{\tiny sh}}$ and $m_{\rm{\tiny SW}}$ are the corresponding Lorentz Factor
and mass of the shocked ISM respectively, and
\begin{equation}\label{a}
m_{\rm{\tiny SW}}=V_{\rm g}m_{\rm p}n(\theta_{j}^{2}/4),
\end{equation}
with $V_{\rm g}$ as the shocked gas volume. Coefficient $\sigma$ differs from 6/17 to 0.73 for
ultra-relativistic and non-relativistic jets (Blandford \& McKee 1976). Wang et al. (2003) took $\sigma\sim$0.7
and we adopt their approximation in our work. Equation (1) and the relativistic kinematic equations
\begin{equation}
(\frac{dR}{dt})_{\rm{a}}=\frac{\beta(\Gamma)c}{1-\beta(\Gamma)\cos\theta};
(\frac{dR}{dt})_{\rm{r}}=\frac{\beta(\Gamma)c}{1+\beta(\Gamma)\cos\theta}
\end{equation}
can be solved and give the relation between the projected angular
separation $\mu$ and time $t$. In equations (2), the subscript `a'
and `r' represent the approaching and receding jets in a pair of
relativistic jets respectively. $R$ is the distance between the jet
and the source, which can be transformed into the angular separation
by $\mu=R\sin\theta/5.3$ kpc, and $\theta$ is the jet inclination
angle to the line of sight. We can get the $\mu-t$ curve numerically
with the above equations. To be consistent with the work done to the
eastern jet, we test first the same initial conditions that
$\Gamma_{0}=3$, $E_{0}=3.6\times10^{44}$ erg,
$\theta=50\textordmasculine$ and $\theta_{j}=1.\textordmasculine5$
as those found by Wang et al. (2003) and would like to find how
these parameters affect the fitting of the jet motions.

In the case of the eastern jet, the number density of the ISM was
assumed by Wang et al. (2003) as a constant in the whole region
outside the central source, thus the interaction was supposed to
have taken place all the way along the jet's path. This assumption
does not work well in the case of its western counterpart, because
the western jet has traveled by 2002 as far as the eastern jet had
traveled by 2000, but was decelerated much more significantly. This
can be seen in Fig. 4. The western jet moved comparatively little
after its appearance and acted like ``hitting into a wall". This
requires much stronger interactions and thus a much denser
environment. We tried to fit the western jet itself but were not
very successful. If we increase the number density, the gas will
simply block the jet and the jet will never travel to such a far
distance (a 10$^{44}$ erg jet can only move to $\sim$7-8 arcsec in 1
cm$^{-3}$ ISM). On the other hand, if the initial energy is too
large ($\sim$10$^{50}$ erg), the jet could penetrate ISM and travel
far but would be very difficult to slow down so quickly (for such
high energy, even a density of 100 cm$^{-3}$ can not bend the curve
down). Changing the initial Lorentz factor has similar results with
changing initial energy and the opening angle of the jet affects the
results quite little. This tendencies are plotted in Fig.4 (a),
where the fitting to the eastern jet shows the model of Wang et al.
(2003) and the corresponding solid-line fitting to the western jet
uses the same model and same parameters. The dashed line and dot
dashed line there represent the results of changing the number
density or the initial kinetic energy to ten times of the eastern
results. All three curves do not match the data points well.

As a result, we consider keeping the initial Lorentz factor and
energy, while changing the geometric assumptions to describe the jet
motions. For the inclination angle, we first examined the paper of
Orosz et al. (2002), where they find a value of 72.6 $\pm$
4.8$\textordmasculine$ from the optical observations. We took this
value, but varying it within their 1-sigma error bars, and find
$\theta=68\textordmasculine$ gives out the best fit. To simplify the
problem, we just fix the inclination angle to this value in our
work.

After setting the inclination angle, we test a model that the ISM
density varies as the distance changes, i.e., the density is lower
in the center and higher in the outside region. For simplicity, we
test the case that the jet traveled first through a ``cavity" with a
constant velocity and then through a dense region and was
decelerated there. For this modification, we introduce another
parameter $r$, the outer radius of the cavity. The ISM number
density is set to be a constant $n$ outside this region and zero
inside. The results are shown in panel (b) and (c) in Fig.4. We
first apply this model to the eastern and the western jet
separately. The results are quite satisfactory but not well
constrained. Because of only 10 data points in constraining 4 free
parameters ($r_{\rm{e}}$, $r_{\rm{w}}$, $n_{\rm{e}}$, $n_{\rm{w}}$),
there are more than one group of parameters that seem work well.
Fitting the two jets simultaneously with a fixed cavity radius helps
a bit. It constrains the fitting tighter and $r=16$ arcsec seems an
acceptable result for the both sides (Fig.4 (b)). However, this is
not a sound assumption. The inferred ISM number density
corresponding to this value is 0.015 cm$^{-3}$ in the east and 0.06
cm$^{-3}$ in the west respectively, indicating a clear asymmetry in
the ISM density on the two side. Thus it is not reasonable to assume
a perfect symmetric geometry. We still need other analysis to help
determining consistent parameters.

\begin{figure}
\includegraphics[width=9cm,,height=18cm]{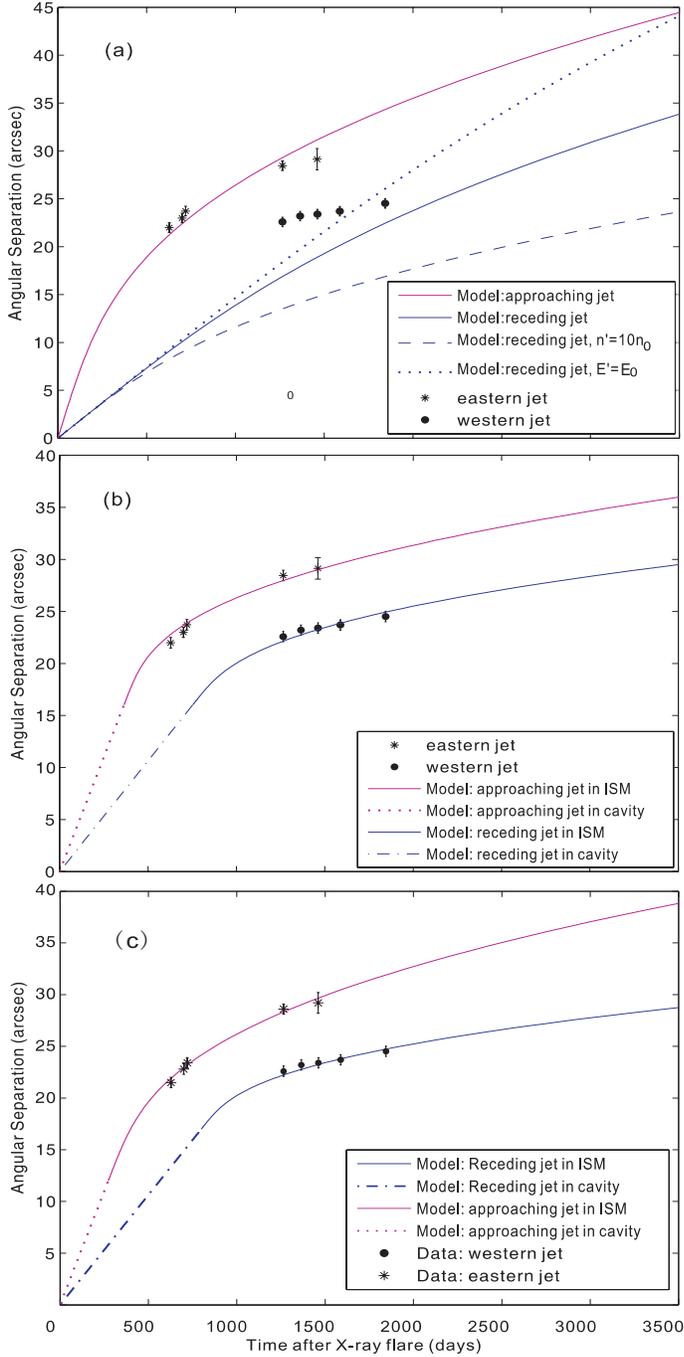}
\caption{Model fittings to the proper motion of the X-ray jets of
XTE J1550-564. In all three panels, the pink lines are the model
predictions of the approaching jet and the blue lines are models
predictions of the receding jet. Panel (a):
$\theta$=50\textordmasculine , $n_{0}=1.5\times10^{-4}$ cm$^{-3}$,
$E_{0}=3.6\times10^{44}$ erg and $r_{\rm{e}}$=$r_{\rm{w}}$=0.
Fitting to the approaching jet is the model from Wang et al. (2003).
The model for the western jet is based on the same parameters. The
dashed and dotted lines are the results if the number density or the
initial kinematic energy is changed to ten times of the eastern
results. Panel (b): model with a symmetric cavity introduced.
Parameters: $\theta$=68\textordmasculine,
$r_{\rm{e}}$=$r_{\rm{w}}$=16 arcsec, $n_{\rm{e}}$=0.015 cm$^{-3}$,
$n_{\rm{w}}$=0.06 cm$^{-3}$. Panel (c): model with an asymmetric
cavity introduced. Parameters: $\theta$=68\textordmasculine,
$r_{\rm{e}}$=12 arcsec, $r_{\rm{w}}$=17 arcsec, $n_{\rm{e}}$=0.0034
cm$^{-3}$, $n_{\rm{w}}$=0.12 cm$^{-3}$.} \label{fig4}
\end{figure}

 \placefigure{fig4}

\subsection{Light Curve Model}
In the standard GRB scenario, the afterglow emission is produced by
the synchrotron radiation or inverse Compton emission of the
accelerated electrons in the shock front of the jets. The forward
shock emission is from the heated electrons in the swept-up ambient
medium, whereas the reverse shock emission is from the electrons of
the jet itself when a shock moves back through the ejecta. The same
processes should also happen in the case of microquasars, with only
the Lorentz factor much lower. In Wang et al. (2003), they tested
these two possibilities and proved that the forward shock emission
would decay too slowly to fit for the observed decay index
$-3.7\pm0.7$ of the eastern jet, because of the continuously heating
effect. However, since only operating once, the reverse shock
emission could decay rather fast as the electrons cooled
adiabatically and this model indeed performed quite well in
describing the data. In the case of the western jet, the energy flux
decayed even more rapidly (decay index $\sim-5.1\pm0.1$, see Fig. 5)
and as a result, we take the reverse shock mechanism in our work.

Assuming the distribution of the electrons obeys a power-law form,
$n({\gamma_{\rm{e}}})d\gamma_{\rm{e}}=K\gamma_{\rm{e}}^{-p}d\gamma_{\rm{e}}$,
for $\gamma_{\textrm{\tiny m}}<\gamma_{e}<\gamma_{\textrm{\tiny
M}}$, the volume emissivity at frequency $\nu'$ in the comoving
frame is given by (Rybicki \& Lightman 1979)
\begin{equation}
j_{\nu'}=\frac{\sqrt{3}q^{3}}{2m_{\rm{e}}c^{2}}(\frac{4{\pi}m_{\rm{e}}c\nu'}{3q})^{\frac{(1-p)}{2}}B_{\perp}^{\frac{(p+1)}{2}}KF_{1}(\nu,\nu'_{\rm{\tiny
m}},\nu'_{\rm{\tiny M}}),
\end{equation}
where
\begin{equation}
F_{1}(\nu,\nu'_{m},\nu'_{M})=\int_{\nu'/\nu'_{M}}^{\nu'/\nu'_{m}}F(x)x^{(p-3)/2}dx,
\end{equation}
with $F(x) = x\int_x^{+\infty}K_{5/3}(t)dt$ and $K_{5/3}(t)$ is the
Bessel function. The physical quantities in these equations include
$q$ and $m_{\rm{e}}$, the charge and mass of the electron,
$B_{\perp}$, the magnetic field strength perpendicular to the
electron velocity, and $\nu'_{m}$ and $\nu'_{M}$, the characteristic
frequencies for electrons with $\gamma_{m}$ and $\gamma_{M}$.

Assuming the reverse shock heats the ejecta at time $t_{0}$ at the
radius $R_{0}$ (with the assumptions of no synchrotron cooling,
conservation of the total number of electrons and the magnetic field
being frozen into the plasma), the physical quantities in the
adiabatically expanding ejecta with radius $R$ evolve as (van der
Laan 1966)
\begin{equation}
\gamma_{m}=\gamma_{m}(t_{0})\frac{R_{0}}{R},
\gamma_{M}=\gamma_{M}(t_{0})\frac{R_{0}}{R},
\end{equation}
\begin{equation}
K=K(t_{0})(\frac{R}{R_{0}})^{-(2+p)}, B_{\perp}=B_{\perp}(t_{0})(\frac{R}{R_{0}})^{-2},
\end{equation}
where the initial values of these quantities are free parameters to
be fitted in the calculation.

With these assumptions, we can then calculate the predicted flux evolution of the jets.
The comoving frequency $\nu'$ relates to our observer frequency $\nu$ by $\nu=D\nu'$,
where $D$ is the Doppler factor and we have $D_{\rm{a}}=1/\Gamma(1-\beta\cos\theta)$
and $D_{\rm{r}}=1/\Gamma(1+\beta\cos\theta)$ for the approaching and receding jets
respectively. Considering the geometry of the emission region, the observed X-ray flux
in 0.3-8 keV band could be estimated by
\begin{equation}
F(\rm{0.3-8
keV})=\int_{\nu_{1}}^{\nu_{2}}[\frac{\theta_{j}^{2}}{4}(\frac{R}{d}){\Delta}RD^{3}j_{\nu'}]d\nu,
\end{equation}
where ${\Delta}R$ is the width of the shock region and is assumed to
be ${\Delta}R=R/10$, after Wang et al. (2003) in the calculation.

To reduce the number of free parameters, we set $\gamma_{m}=100$ in
our calculation because the results are quite insensitive to this
value. According to our kinematic model in section 4.1, we choose
the time that the reverse shock takes place to be the time that the
Lorentz factor reduced to $1/\sqrt{2}$ of its original value, which
is called deceleration timescale $t_{\rm{dec}}$ in the GRB external
shock model, which is supposed to be the strongest point of the
external shock (Private communication with X. Y. Wang). Then we fit
the data to find out the initial values of $K$ and $B_{\perp}$.

Just like the fitting to the kinematics of the jets, we could not
decide the best fitting result only using the flux data since we
could always find one group of parameters that fit the flux data
approximately well for each parameter $r$. Thus, we combine the
kinematic and light curve fitting together to find some more useful
hints.

We know that the energy and the number density of the gas in the
pre-shock and post-shock regions are connected by the jump
conditions $n'=\zeta(\Gamma)n$ and $e'=\eta(\Gamma)nm_{p}c^{2}$,
where $\zeta(\Gamma)$ and $\eta(\Gamma)$ are coefficients related to
the jet velocity (Wang et al. 2003 ). Therefore if we assume the
shocked electrons and the magnetic field acquire constant fractions
($\epsilon_{\rm{e}}$ and $\epsilon_{B}$) of the total shock energy,
we have
\begin{equation}
\gamma_{m}=\epsilon_{e}\frac{p-2}{p-1}\frac{m_{p}}{m_{e}}(\Gamma-1),
K=(p-1)n'\gamma_{m}^{p-1},
\end{equation}
and
\begin{equation}
B_{\perp}=\sqrt{8\pi\epsilon_{B}e'}
\end{equation}
for p$>$2. Since we have assumed that $p$ and $\Gamma_{0}$ (so that
$1/\sqrt{2} \Gamma_{0}$) are equal for the two jets, if we further
assume that factor $\epsilon_{e}$ of the eastern and the western
jets is also the same, we may infer that $K\propto{e'}\propto{n}$
for the two jets. We therefore search for the combination of
parameters that could satisfy the kinematic and light curve fitting,
as well as the relationship
$K_{\rm{e}}/K_{\rm{w}}{\sim}n_{\rm{e}}/n_{\rm{w}}$.

To search for the best parameters fitting the data, we follow the
following procedure. With numerical calculations, for the 19 data
points (including 10 kinematic data and 9 light curve data) we build
a large 8-dimensional database for the eight parameters to be
estimated, with one constraining relationship,
$K_{\rm{e}}/K_{\rm{w}}{\sim}n_{\rm{e}}/n_{\rm{w}}$. Intuitively it
looks very difficult to determine so many parameters from so few
data points. However, one approach we take is that we divide the
fittings into two stages: we first fit the kinematic data regardless
of the lightcurve information and get a series of $(r, n)$ all of
which describe the data almost equally well. Then, for each set of
$(r, n)$, we fit the lightcurve and calculate the joint $\chi^{2}$
value to choose the group of parameters yielding the least
$\chi^{2}$ value. The advantage  of this approach is that, since
$(r, n)$ is fixed every time in the fitting of the lightcurve, we
consider the kinematic fittings to be independent of the lightcurve
fittings in calculating $\chi^{2}$. At each search step, the total
$\chi^{2}$ is calculated corresponding to the 19 data points.

A set of parameters that yields the minimum total $\chi^{2}$ is
taken as our best fitting parameters. The resulting total $\chi^{2}$
is 7.05 and the errors are estimated by searching for the range for
one parameter that can change the $\chi^{2}$ value by a given delta
while keeping all the other parameters fixed (Press et al. 1992,
``Constant Chi-Squared Boundaries as Confidence Limits"). The result
corresponding to the lightcurve fitting is shown in Fig.5 and the
corresponding kinematic fitting is shown in panel (c) in Fig.4. The
best fitting parameters are listed in Table 4. From the parameters,
we conclude that the boundary of the cavity lies at $r\sim$12 arcsec
to the east and $\sim$17 arcsec to the west of the central source.
The corresponding number density of the ISM outside this boundary is
$\sim$0.0034 cm$^{-3}$ and $\sim$0.12 cm$^{-3}$, respectively. These
values are both lower than the canonical ISM value of $\sim$1
cm$^{-3}$, although the value in the western region is much higher
than in the eastern region. The asymmetry of the density on the both
sides probably involves the generation history of the cavity, which
should be explored with further studies. The electron energy
fraction relationship is satisfied as
$K_{\rm{e}}/K_{\rm{w}}{\sim}n_{\rm{e}}/n_{\rm{w}}\sim0.03$. However,
we should mention here that the other relation concerning the
magnetic field strength (if $\epsilon_{\tiny{B}}$ are equal on the
two sides, then
$B_{\rm{e}}/B_{\rm{w}}{\propto}\sqrt{e'_{\rm{e}}/e'_{\rm{w}}}{\propto}\sqrt{n_{\rm{e}}/n_{\rm{w}}}$)
could not be satisfied simultaneously by these parameters. Although
the cavity radius and the number density are allowed to vary
significantly, the best fitted magnetic field strength remains quite
stable ($\sim$0.35-0.55 mG). One possible interpretation for this is
that the equipartition parameter varies as the physical conditions
of the jet varies or equipartition assumptions do not hold here for
the magnetic fields since this is not a steady system; an
alternative explanation may involve the {\it in situ} generation (or
amplification) of the magnetic field (e.g., originated from external
ISM instead of from the jet itself).

In the fitting, the assumed initial Lorentz factor is
$\Gamma_{0}=3$, indicating an initial velocity of $v\sim0.943c$. The
fixed spectral index, $p=2.2$, is consistent with the value obtained
by Xue et al. (2008) in their broadband spectral analysis
($p\sim2.20-2.31$). The fitted magnetic field strengths is lower
than the value given by Xue et al. (2008) ($B\sim1-32$ mG). However,
they claimed a large uncertainty in their fittings. We look forward
to further analysis to explore this more clearly and take our
fitting as a reasonable one at this stage.

\placetable{table4}

\begin{deluxetable}{rrrrrr}
\tabletypesize{\footnotesize} \tablecolumns{6} \tablewidth{0pc}
\tablecaption{Model Fitting Results of the Eastern and Western
Jet\label{table4}} \tablehead{\colhead{} & \colhead{Eastern jet}&
\colhead{Western jet}& \colhead{$n_{\rm{e}}/n_{\rm{w}}$} &
\colhead{$K_{\rm{e}}/K_{\rm{w}}$} &\colhead{Comments}
 }

 \startdata
$r$ (arcsec)  &$12\pm0.3$   &$17\pm0.1$  &   &  &Cavity radius \\
$n$ (cm$^{-3}$)  &$(3.4\pm0.3)\times10^{-3}$   &$0.12\pm0.01$  &   &  &ISM density outside the cavity \\
$\theta$ (\textordmasculine)   &68   &68  &   &   &Line of sight angle \\
$r_{0}$ (arcsec)  &15.0  &17.9  &0.03   &0.03  &Reverse shock hitting place\tablenotemark{a} \\
$t_{0}$ (days)  &343   &842   &   &   &Reverse shock hitting time\tablenotemark{a} \\
$B_{\perp}(t_{0})$ (mG)  &$0.51\pm0.03$  &$0.38\pm0.003$   &   &   &Initial value at $t_{0}$ \\
$K(t_{0})$ (cm$^{-3})$   &$(9.0\pm1)\times10^{-3}$   &$0.32\pm0.03$   &  &   &Initial value at $t_{0}$\\
$\chi^{2}$ for fitting &kinematics: 1.46 &kinematics: 1.55 & & &total $\chi^{2}$=7.05 reaches the minimum \\
           &light curve: 1.68 &light curve: 1.68 & & & \\
 \enddata \tablenotetext{a}{The place and time
that the Lorentz factor reduced to $1/\sqrt{2}$ of its original
value, according to the kinematic model.}

\end{deluxetable}

\begin{figure}
\includegraphics[width=12cm]{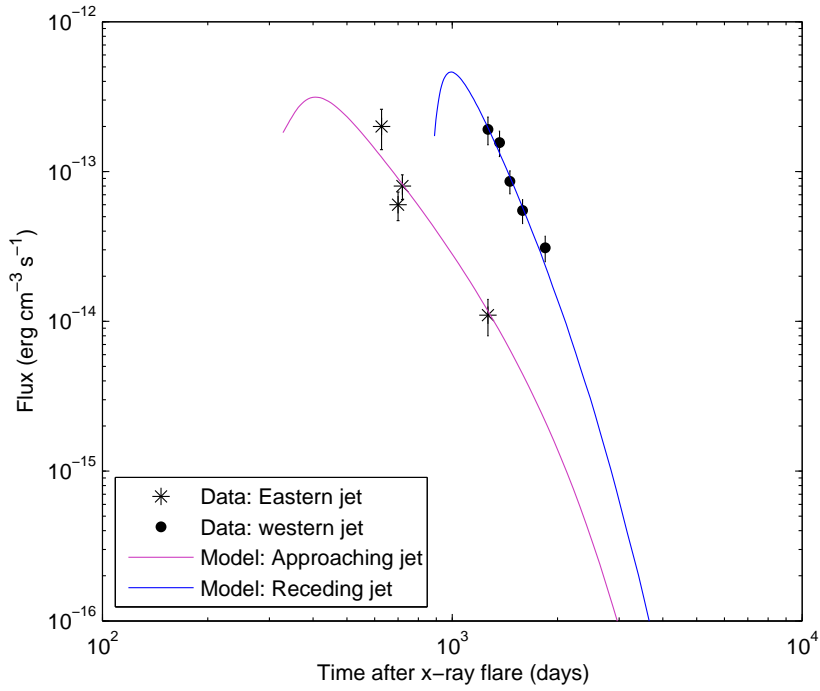}
\caption{Model fittings to the X-ray light curves of the eastern and
western jets. A power-law plus Galactic absorption spectral model is
used to obtain the energy flux in the 0.3-8 keV band. The two solid
lines are the theoretical model fittings for the reverse shock
heated ejecta emission.} \label{fig5}
\end{figure}

\placefigure{fig5}

\section{ANALYSIS OF JETS IN H1743-322}
H 1743-322 is an X-ray transient first discovered in 1977
(Kaluzienski \& Holt 1977). It was then classified as a black hole
candidate in 1984 based on its X-ray spectral characteristics (White
\& Marshall 1984). New activity was found by INTEGRAL in 2003
(Revnivtsev et al. 2003), and a bright radio flare was observed by
VLA on 2003 April 8 (Rupen et al. 2003). $\it Chandra$ X-ray and
ATCA radio observations from 2003 November to 2004 June revealed the
presence of large-scale jets on both sides (Rupen et al. 2004;
Corbel et al. 2005). The source is then labeled as a ``microquasar".

We take the data from the work of Corbel et al. (2005) and fit the
proper motion with the external shock model. Following their
approach, the ejection date is set to the time of the major radio
flare (2003 April 8). The distance to H 1743-322 is basically
unknown now. However, its location toward the Galactic bulge could
possibly imply a Galactic center location. Assuming a source
distance of 8 kpc (distance to the Galactic center), the first radio
detection of the jets gives out an intrinsic velocity of the
ejection of $\beta=v/c=0.79$ and an angle of
$\theta=73\textordmasculine$ for the axis of the jets (Corbel et al.
2005).

However, the evolution of the jets shows that the velocity is not
constant all the way through. Deceleration is required since the
linear extrapolation of the proper motion data yields an ejection
date earlier than the chosen zero point time, as shown in panel (a)
in Fig. 6. The linear extrapolation of the zero point and the first
data point also gives out lines with steeper slope than the real
data.

As a result, we also apply the external shock model to this source.
The line of sight angle is set to be $\theta=73\textordmasculine$ in
all fittings. It is worth mentioning that the 73 degree angle is
only valid if the jets have both been propagating outwards with the
same speed and the same angle to the line of sight and the source is
assumed to be located at the Galactic center with a distance of 8
kpc. However, since there are no better constraints on the jet
orientation, we simply take this value as a reasonable
approximation. We first test the model that assumes the source
located in a continual gas medium. In this case, the jets would
decelerate gradually after ejection. The model fits the data quite
well this time (panel (b) in Fig.6, with $\chi^{2}$=1.03 for the
approaching jet and $\chi^{2}$=0.98 for the receding jet, with
degree of freedom=2 for each case), indicating that a cavity is not
required in this source. However, we also notice that the best
fitting density of the environment ISM is $n\sim3\times10^{-4}$
cm$^{-3}$, much lower than the canonical Galactic ISM value. We also
apply the cavity model to the fitting, which describes the data
equally well (panel (c) in Fig. 6, with $\chi^{2}$=0.95 for the
approaching and $\chi^{2}$=0.97 for the receding jet, with degree of
freedom=2 for each case), with a cavity region with radius of 3
arcsec on both sides and the gas density of $n\sim3\times10^{-3}$
cm$^{-3}$. This value is one magnitude higher than the previous
attempt, but still quite low. Due to the limited data, we cannot
test if the environment outside the central source is symmetric or
not. The flux is not fitted for this source because of the limited
number of X-ray data points.

\begin{figure}
\includegraphics[width=9cm,height=18cm]{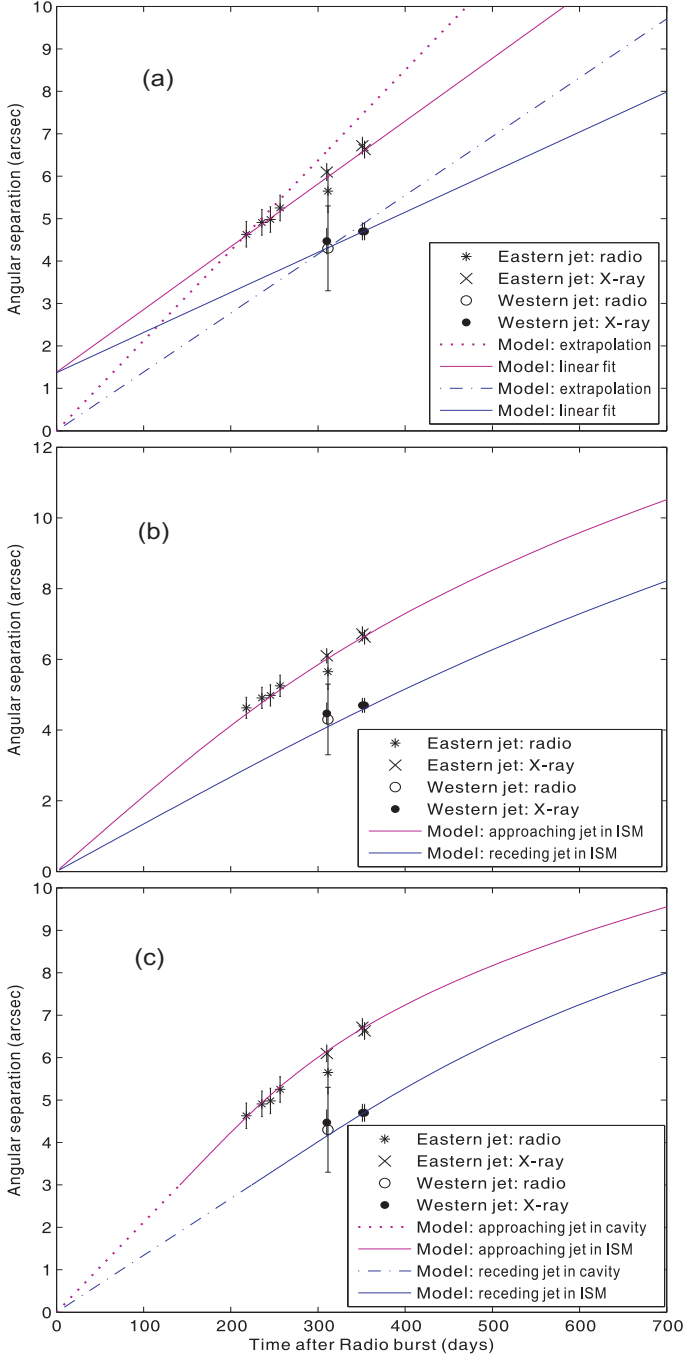}
\caption{Model fittings to the proper motion of the radio and X-ray
jets of H 1743-322. Panel (a) shows the constant velocity fitting
and Panel (b) and (c) are deceleration model fittings. Panel (a):
Dotted lines: extrapolation of the zero point time and the first
radio detection. Solid lines: linear fittings to the data set with
constant velocity. Panel (b): Deceleration in constant density
medium. Results: $\Gamma_{0}=1.65, \theta=73\textordmasculine,
E_{0}=1\times10^{44}$ erg, $n\sim3\times10^{-4}$ cm$^{-3}$. Panel
(c): Deceleration in medium outside a cavity region. Results:
$\Gamma_{0}=1.65, \theta=73\textordmasculine, E_{0}=1\times10^{44}$
erg, $r_{\rm{e}}=r_{\rm{w}}=3$ arcsec, $n\sim3\times10^{-3}$
cm$^{-3}$. } \label{fig6}
\end{figure}

\placefigure{fig6}

\section{GENERAL PICTURE OF MICROQUASAR ENVIRONMENT}
The above analyses of XTE J1550-564 and H 1743-322 led us to the
conclusion that in microquasars the interactions between the ejecta
and the environmental gas play major roles in the jet evolution. We
could further argue that the low density of the environment is a
necessary requirement for the jet to develop to a long distance.

Heinz (2002) derived the scale relations for the jets from accreting
black holes, using a simple analytic model. He adopted the jet-ISM
interaction scenario to estimate the slow-down distance of the
ejecta:
\begin{equation}
d_{\rm{slow}}\sim10^{16}\rm{cm}(E_{44}/\Gamma_{5}^{2}n_{x}\theta_{5}^{2})^{1/3},
\end{equation}
where $E_{\rm{kin}}\equiv10^{44}E_{44}$ erg is the kinetic energy,
$\Gamma_{5}\equiv\Gamma/5$ is the Lorentz factor of the jet, $n_{x}$
is the external gas density in the unit of cm$^{-3}$ and
$\theta=5\textordmasculine\theta_{5}$ is the opening angle of the
jet. He applied this estimation to microquasars GRS 1915+105 and GRO
J1655-40. In both cases, he found that the upper limit on the gas
density is roughly $10^{-3}$ cm$^{-3}$, given $d_{\rm{slow}}$ of
~0.05pc, quite consistent with the ISM density found for XTE
J1550-564 and H1743-322 in this work. For a larger $d_{\rm{slow}}$,
a lower $n_{x}$ should be required.

\begin{figure}
\includegraphics[width=12cm]{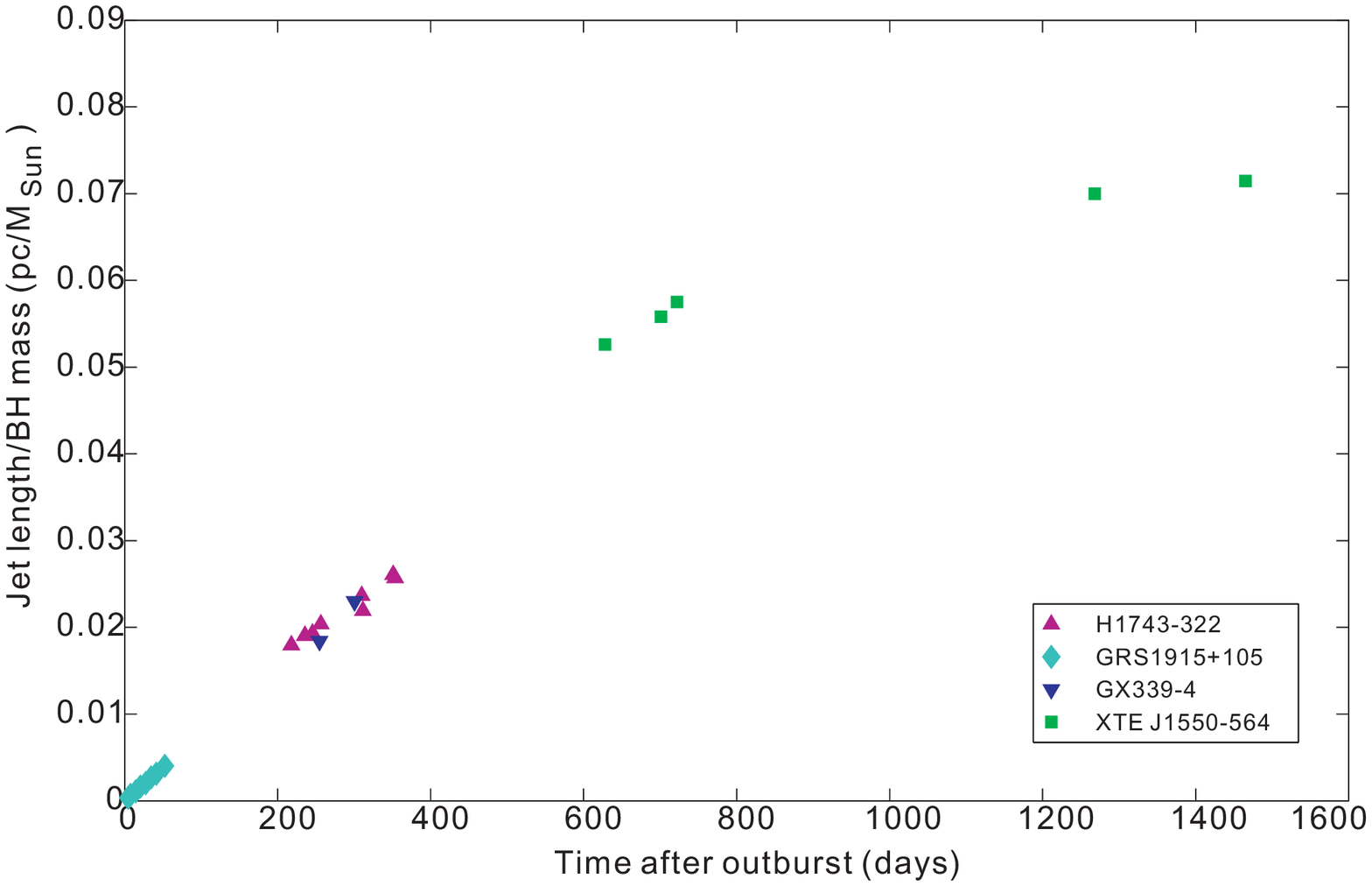}
\caption{Normalized proper motions of microquasars' approaching
jets, obtained by dividing the actual jet length by the central
compact object mass. For the mass and distance estimations, please
refer to the text, see section 6. Jet kinematics data are taken from
Rodriguez \& Mirabel (1999), Gallo et al. (2004), Corbel et al.
(2005) and this paper respectively.} \label{fig7}
\end{figure}

\placefigure{fig7}

Heinz's approach led to an interesting comparison between the
microquasar jets and the radio quasar jets. He claimed that the jets
length in GRS 1915+105 ($l>$0.04 pc) would correspond to a jet
length of 4 Mpc when scaled by $M$ (mass of the central black hole)
to AGN conditions (such as M87 or Cyg A) (2002). We could follow
this way and infer that the jet in XTE J1550-564 ($l>$0.5 pc) will
correspond to a 50 Mpc long AGN jet, which has never been detected.
This again confirms that the environment of microquasars should be
comparatively vacuous, in a dynamical sense even (although not in
the absolute sense) less dense than the AGN environment (the
intergalactic medium of densities between 10$^{-5}$ cm$^{-3}$ to
10$^{-2}$ cm$^{-3}$), when compared to the thrust of the jets in
microquasars and AGNs. When presented in unit of pc with the
information of source distance and scaled by the central black hole
mass, the proper motions of different microquasar jets can be
plotted in one figure (Fig. 7) (GRS 1915+105: 14$\pm$4 $M_{\odot}$
(Greiner et al. 2001), 12.5 kpc (Rodr\'{\i}guez et al. 1995); GX
339-4: 5.8$\pm$0.5 $M_{\odot}$ (Hynes et al. 2003), lower limit
estimation of distance of 4 kpc (Zdziarski et al. 1998); XTE
J1550-564: 10.5$\pm$1.0 $M_{\odot}$ and 5.3 kpc (Orosz et al. 2002);
H 1743-322: since no good mass and distance observation up to date,
we follow the assumption of 10 $M_{\odot}$ (Miller et al. 2006) and
galactic center origin of 8 kpc (Corbel et al. 2005)). It is
interesting to notice from this figure that although the detailed
properties of the binary systems (e.g., the compact objects and the
companion stars) differ a lot from source to source, the normalized
jet proper motions are quite consistent. This is not surprising
since the jet properties are set by the compact object (and its
immediate neighborhood) and their evolutions are mostly influenced
by the surrounding environment. Although the companion star provides
the material which is eventually launched into the jets, and thus
affects the mass transfer rate and the mass flux in the jets, it
nevertheless does not seem to play significant roles for the
structure and evolution of the large scale jets and cavities
surrounding the microquasar. Therefore the nature of accretion disk
physics dominates over the nurture of the companion star, as far as
the jets are concerned. Thus it leads us to the attempt to
generalize the microquasar jets in one consistent picture.

This figure, together with other observational data concerning the
motionless large scale jets (e.g., Cyg X-1, SS433), has led us to
the suggestion that all microquasar jets can be classified into
roughly three groups: small scale moving jets, large scale moving
jets and large scale jet relics. For the first type, the ``small
jets", only radio emissions are detected. The jets are always
relatively close to the central source and dissipate very quickly.
Examples of this class include GRS 1915+105 (Rodr\'{\i}guez \&
Mirabel 1999; Miller-Jones et al. 2007), GRO J1655-40 (Hjellming \&
Rupen 1995), and Cyg X-3 (Mart\'{\i} et al. 2001). The typical
spatial scale is 0$\sim$0.05 pc and the time scale is several tens
of days. In this class, no obvious deceleration is observed before
the jets become too faint. For the second type, the ``large jets",
both X-ray and radio detections are obtained, at a place far from
the central source several years after the outburst. Examples are
XTE J1550-564, H1743-322, and GX 339-4 (Gallo et al. 2004). The
typical jet traveling distance for this type is 0.2$\sim$0.5 pc from
the central engine. Deceleration is clearly observed in this type of
sources. The last type, the ``large relics", is a kind of diffuse
structures observed in radio, optical and X-ray band, often ring or
round nebula shaped that are not moving at all. In this class, some well studied sources, Cygnus X-1 (Gallo et al. 2005), SS433 (Dubner el
al. 1998), Circinus X-1 (Stewart et al. 1993, Tudose et al. 2006)
and GRS 1915+105 (Kaiser et al. 2004) are included. The typical
scale for this kind is 1$\sim$30 pc, an order of magnitude larger
than the second type. The estimated lifetime often exceeds one
million years, indicating that they are related to previous
outbursts.

From these properties, it is reasonable to further suggest a
consistent picture involving all the sources together. We make a
conjecture that extreme low density regions, or namely, large scale
cavities, exist in all microquasar systems. The ``small jets"
observed right after the ejection are just traveling through these
cavities. Since there are few or none interactions between the jets
and the surrounding gas in this region, the jets travel without
obvious deceleration. The emission mechanism is synchrotron
radiation by particles accelerated in the initial outburst. The
emissions of jets decay very quickly and are not detectable after
several tens of days. In some cases (e.g. XTE J1550-564), the cavity
has a dense (compared to the cavity) boundary at some radius and the
interactions between the jets and the boundary gas heat the
particles again and thus make the jets detectable again. Those are
the ``large jets". The emission mechanism then is synchrotron
radiation by the re-heated particles in the external shocks. Then,
after these interactions, the jets lost most of their kinematic
energy into the ISM gradually, causing the latter to expand to large
scale structures, the ``large relics", in a comparatively long time
(thousands to several millions of years) and create the observed
nebula-like jet-inflated bubbles.

Different systems are allowed to fall into more than one of the
three postulated groups. For example, Cygnus X-1 has small-scale
ejecta (Fender et al., 2006) and also has a large-scale relic (Gallo
et al., 2005). Also XTE J1550-564 has small-scale ejecta
(Hannikainen et al., 2001) and also large-scale jets (Corbel et al.,
2002). As we proposed in the above paragraph, the three
observational groups are perhaps just three stages in a complete
ejection evolution process: the ``small jets" observed in the
outburst events will travel in dark and manifest themselves as the
``large jets" when they hit the boundary of the cavity and
eventually develop into the ``large relics" gradually by the
continuous interactions and energy dissipation. If this conjecture
holds, then the three groups may not be actually ``three" but just
three phenomena in ``one" system. The three groups identified here
are probably just observational manifestations of one generic
mechanism, and different sources can belong to multiple groups.

If this general model of microquasars is true, properties of
different jet systems could be dealt with separately in one
consistent scheme. One implication is that the electrons from the
``large jet" and the ``small jets" belong to different groups of
electrons, from the ejecta itself and the ISM respectively, which
means we could not simply use the properties of electron
distributions in the ``large jet" blobs to determine the emission
properties of the central source in the ejection (Xue et al. 2008).

One important remaining question would be why there are such vacuous
regions and how to generate them. There are several possibilities,
involving previous outbursts, jets or winds activities. Before we go
further into these possibilities, we should first distinguish one
class from the others. We have used the term ``cavity" all through
our discussion, but actually there are two geometric possibilities
for a cavity. One is a ``cavity", the other is a ``tunnel". Instead
of a spherical cavity, it is possible that there is only a vacuum
conical path outside the central engine. It may be created by
previous jet ejection events and got fully developed through a
series of ejections. This could explain the collimation of the jets
naturally, but it requires almost continuous jet ejections (contrary
to the episodic jet ejections discussed here) to prevent the
surrounding gas to fall back and fill up the thin path between
ejections.

This assumption is of great interest because it provides a possible
connection between the large-scale structures and the persistent,
low-power, steady jets reported in several important observations
(etc. Cyg X-1 (Stirling et al. 2001) and GRS 1915+105 (Dhawan et al.
2000)). Kaiser and Alexander (1997, hereafter KA97) has developed a
model for radio galaxies in which continuous and collimated jet may
clear its way out and create a hot spot at the end of its path, and
the hot electrons escaped from the shocks would form a ``cocoon"
outside the central compact object. If this is also the case for
Galactic accretion systems, as postulated for Cygnus X-1 (Gallo et
al., 2005), Circinus X-1 (Tudose et al., 2006), and GRS 1915+105
(Kaiser et al., 2004), then the vacuous regions (tunnel and cocoon)
and the ``large relics" are perhaps both the results of the
continuous jets traveling in dark. The two Galactic center sources
1E1740-294 and GRS 1958-258 may also belong to this case. In this
frame of assumptions, the episodic outbursts are not included here:
the outbursts will increase the instantaneous mass flux down the
jet, however since low-mass X-ray binaries spend the majority of
their time in a low-luminosity quiescent state, the outbursts will
have little effect on the time-averaged jet luminosity, and thus on
the long-term evolution of the relics. However, we may infer that if
there are dark continuous jets and dark relics in a system like XTE
J1550-564, a strong episodic outburst would just hit and brighten
the boundary of the relic and may provide a bridge between the
source and the relic. If there are multiple outbursts, we will
expect repeated ``large jets" in the same direction with larger and
larger lengths. Future observations would then be able to testify
this possibility.

\placetable{table5}

\begin{deluxetable}{rrrrrrrr}
\tabletypesize{\footnotesize} \tablecolumns{8} \tablewidth{0pc}
\tablecaption{Wind properties of microquasars\label{table5}}
\tablehead{\colhead{Source Name} & \colhead{Equip.}& \colhead{M
($M_{\odot}$)}& \colhead{$r_{\rm g}$ (km)\tablenotemark{1}} &
\colhead{Velocity (km/s)} &\colhead{Line width (km/s)}
&\colhead{$R_{\rm{\rm eje}}/r_{\rm g}$} &\colhead{$V_{\rm{esc}}$
(km/s)\tablenotemark{2}}  }

 \startdata
GRO J1655-40  &$\it Chandra$\tablenotemark{3}   &7  &10.5   &300-1000  &300-500 &$10^{4.7}$ &1900 \\
  &            &  &  &           & &$10^{5.7}$  &600\\
  &XMM-Newton\tablenotemark{4}  &  &  &2600-4500  & &$5\times10^{3}$ &6000\\
  &            &  &  &           & &$2\times10^{4}$  &3000\\
H 1743-322 &$\it Chandra$ \& RXTE\tablenotemark{5} &10 &15 &$700\pm200$ &$1800\pm400$ &$10^{4}$ &4240\\
GRS 1915+105 &ASCA\tablenotemark{6} &$14\pm4$ &21 &1000 & &$10^{5}$ &1340\\
  &    Chandra\tablenotemark{7}        &  &             &$1100^{+360}_{-300}$ &$980^{+450}_{-420}$  &$3\times10^{5}$ &770\\
 \enddata
\tablecomments{\textbf{1}. $r_{\rm g}=GM_{\rm BH}/c^{2}; \textbf{2}.
V_{\rm{esc}}(r)=\sqrt{2GM_{\rm BH}/r}$; \textbf{3}. Netzer, 2006;
\textbf{4}. Sala et al. 2006; \textbf{5}. Miller et al. 2006;
\textbf{6}. Kotani et al. 2000; \textbf{7}. Neilsen \& Lee 2009;}

\end{deluxetable}

Apart from this case, several other possibilities may take place.
One possible way to create the observed cavities is through recent
supernovae outbursts that produced the central black holes. However,
only HMXBs should be close to their parent supernova remnants (being
young); accumulated velocity dispersion (scattering from spiral arms
or giant molecular clouds over time) will carry LMXBs far from their
natal supernova remnant in a few Galactic orbits, e.g. for GRO
J1655-40 (Israelian et al. 1999). Furthermore, some sources such as
Cygnus X-1 most likely never had supernovae, because of their low
peculiar velocity (Mirabel \& Rodr\'{\i}gues 2003). In the case of
recent supernovae ($<10^{5}$ years) and low kick-out velocity for
the black hole, SNRs surrounding them should have been detected, and
for much older supernovae and very low kick-out velocity for the
black hole, the microquasars may still resides in the SNRs but
regular interstellar medium should have filled those very old SNRs
now.

Winds from progenitors of the central compact objects at their last
stages may also be an interpretation. Cavities in hydrogen maps have
been detected around neutron stars and were supposed to be driven by
wind from the progenitors of the stars (Gaensler et al. 2005).
However, the lifetime of such bubbles were estimated to be of order
of $\sim$1 Myrs. For a region still as vacuous as we have estimated
in our studies, the compact object is expected to be much younger
than that. Winds of the companion star in the binary are also a
possible candidate. Wind-blown bubbles and shells are observed
around evolved massive stars, with typical radii of $\sim$2-10 pc
for Wolf-Rayet stars (Gruendl et al. 2000) and $\sim$0.1-2.3 pc for
luminous blue supergiants (Smith et al. 2007). However, the
estimated density at the bubble shells is of order of $\sim$100
cm$^{-3}$, simply too high for our case, and among all the well
established microquasar systems, only the companion star in Cyg X-1
is O type (O9.7Iab) (Remillard \& McClintock 2006), whereas most of
the other systems contain only M, K or G type companion stars, which
are not powerful enough to generate such large scale bubbles.

A plausible possibility involves the accretion disk winds. In this
scenario, it is the disk winds with mild or non-relativistic
velocities that have pushed away the surrounding ISM gas and created
the low density regions surrounding the central source. The wind
should go as shells that travel in a ballistic way while interacting
with the ISM cold gas. Indeed, in several microquasars, fast ionized
winds have already been observed (Miller et al. 2006; Kotani et al.
2000; Netzer, 2006; Sala et al. 2006; Fuchs et al. 2006). Strong
absorption metal lines are detected with obvious blueshift. The
speed of the outflow could be calculated from the blueshift of these
lines. The ejection place could also be estimated using the measured
ionization parameters. We list some of the results in Table 5,
including the calculated escape velocity at the inferred ejection
radius ($r_{\rm g}=GM_{\rm BH}/c^{2}$ and
$v_{\rm{esc}}(r)=\sqrt{2GM_{\rm BH}/r}=\sqrt{2r_{\rm g}c^{2}/r}$).
Also, in another interesting work, Fuchs et al. (2006) showed that
there are indeed strong subrelativistic winds coming out of the SS
433 system, very much alike the winds from a WR star or from a thick
torus or envelope outside the accretion disk and the central object.
The estimated mass lost rate is $4.7-7.3\times10^{-5} M_{\odot}$
yr$^{-1}$ for discontinuous winds and 3 times higher for continuous
winds, much higher than the observed jet mass loss of $\sim$10$^{-7}
M_{\odot}$ yr$^{-1}$ in this system, and a double-cone structure may
be formed by these winds. These inferences are also supported by
Blundell et al. (2001), where the equatorial radio emission seen in
SS433 was taken as evidence for a disc wind. These observations
support the idea of co-existence of jets and winds in the
microquasars systems. We also notice that the wind velocity in
H1743-322 is lower than those in GRS 1915 +105 and GRO J1655-40,
suggesting larger cavities are formed for the latter two sources
than for the former one. It is consistent with the fact that no
decelerating jets have been observed in these two sources up to now.

However, whether the winds are powerful enough to create such large
cavities in all cases for all sources is unclear at this stage. The
terminal places of the wind shells (which determine the size of the
cavity) are determined by the wind energy and geometry,
specifically, the wind velocity at the ejection place and the
opening angle, and the interaction conditions between winds and the
gas. These parameters are still very uncertain nowadays because of
the lack of enough studies and the densities of the winds are only
estimated at the foot of the ejection place. We would roughly
estimate that, in order for the wind to move to large radii far from
the black hole, the wind velocity should be quite close to the
escape velocity at the ejection radius. From table 5, we can see
that most of the observed wind velocities are comparable to,
although not as large as, the local escape velocity, with the
exception of the recently discovered disk wind of about its escape
velocity in GRS 1915+105 in its soft state, with mass loss rate
sufficient to suppress jet ejection (Neilsen \& Lee 2009). Therefore
in the last case the accretion disk wind alone (because no jet is
expected in the soft state) may play a significant role for forming
a large scale cavity. However for other cases the winds alone may
not be enough to power the cavities, due to insufficient velocities.

Therefore some kinds of acceleration or velocity-maintaining
mechanisms may be required, in order for the winds to expand far
enough ($\sim$0.1 pc to 10 pc). Magnetic fields may play an
important role here. It has been known that rotating wind-up
magnetic fields can accelerate and launch jets (see Spruit 2008 and
references therein); therefore continuous high velocity winds can
set up the favorable condition for producing continuous outflows or
jets with velocities around or even exceeding the escaping
velocities of these systems. Once continuous high velocity outflows
or jets are produced, they should be able to inflate their
surrounding interstellar medium to form bubbles, as we have
discussed earlier within the framework of the KA97 model.

Finally, we comment that it seems counter intuitive that the boundary of the cavity lies further in the western
side, yet the ISM density is also higher in the west. It would be a problem if the cavity is produced by the
observed large scale two-sided jets with equal power. Therefore this asymmetry may be used to argue against such
a model. However, the asymmetric cavity may be naturally produced if the system has a non-negligible space
velocity with respect to the local ISM along the eastern direction. In this case additional energy is injected
into the ISM on the head-on side along the direction of motion of the system, in analogy to process producing
the head-on bow-shock of young spin-down powered pulsars moving in ISM (e.g., Caraveo et al 2003). Therefore the
cavity on the approaching side should be emptier (with smaller density) than the opposite wind or jet, unless
the system moves perpendicular to the wind or jet. The system itself is also catching up with the cavity on the
approaching side, placing itself closer to the boundary of the cavity on the approaching side. For example, a
space velocity of 100 km/s with respect to the local ISM would displace the source by 0.1 pc in less than 1000
years. In the absence of data for this system on its proper motion and the power of wind or continuous (dark)
jets, we will not speculate further on what combination of parameters may eventually produce the observed
asymmetry of the cavity. This should be tested in the future. Heinz et al. (2008) have proposed that the
microquasar's space velocity should be important and leave trails behind their moving path in the Galaxy. They
estimated the volume of radio plasma released by these microquasars and claimed that they should be observable
at low frequencies. Future observational confirmation of their prediction for XTE J1550-564 would provide
evidence for our conjecture.

\section{CONCLUSIONS AND PERSPECTIVES}
Two large-scale X-ray jets have been observed in XTE J1550-564. We
have analyzed the $\it Chandra$ X-ray data for these two jets and
fitted their kinematics and light curves with the external shock
model in the GRB afterglow model. In this model, the interactions
between the jet material and the surrounding ISM slow the jets down
and accelerate the jet particles to radiate the observed radio and
X-ray synchrotron radiations. Under such assumptions, the number
density of the ISM in the surrounding regions of the central black
hole is a key parameter. We found in the fitting that this number
density has to be extremely low, or the strong interactions would
block the jet's expansion. However on the other hand, the number
density at $\sim$20 arcsec has to be comparatively high (but still
lower than the typical ISM) in order to allow the jet to be
decelerated quickly, as the observational data indicated. Thus, a
model consisting of two different regions, a cavity and its gas
boundaries, is tested and shown to work well. The cavity is found to
be large scale ($\sim$0.4 pc in size) and asymmetric ($n_{\rm
e}/n_{\rm w}\sim0.03$). The reason of the asymmetry of the
environment on the two sides is not clear from current observational
data. The deceleration described in the model is quite obvious (from
$\Gamma_{0}=3$ to $\Gamma_{t}\simeq1$ at $t\sim1500$ days), thus it
is the first time that we observe the whole process of the
deceleration of a pair of relativistic jets.

The similar analysis of H1743-322 also supports this jet-ISM
interaction scenario. Large scale decelerating jets are also
observed in this source and the interaction model describes the data
consistently. Whether a cavity exists is not clear in this case;
nevertheless the density is  also found to be very low, compared to
the canonical Galactic value.

With these studies, a generic model of the environment of
microquasars is proposed. We suggest that microquasars are located
in large scale and very low density cavities. The cavity provides
the space for the jets to go through freely and the interactions
between the jets and the material outside the cavity provide the
deceleration mechanism of the jets. The cavity is likely created by
continuous high velocity outflows or jets, which are seeded by
accretion disk winds that have been observed in several microquasar
systems producing strong absorption lines.

Microquasars are powerful probes of both the central engine and
their surrounding environment. More studies of the jets behaviors
may give us information on the ISM gas properties, as well as the
ejecta components. The link between jets and wind is a key issue in
current studies of accreting systems. If the general scenario we
have proposed here is further confirmed, we then need to understand
the mechanism producing these outflows (winds and jets) and the
connections between them. Detailed analysis of the geometry and
timing properties of the ejecta in the future would be of great
significance; it will not only provide insights of the jet formation
process, but also offer another approach into black hole physics and
accretion flow dynamics.

\acknowledgments

\textbf{Acknowledgments}

We are grateful to X.Y. Wang for many helps in this work. We also
thank Z.G. Dai and R. Soria for useful discussions. We appreciate
very much the insightful comments and helpful suggestions by the
anonymous referee. SNZ acknowledges partial funding support by the
Yangtze Endowment from the Ministry of Education at Tsinghua
University, Directional Research Project of the Chinese Academy of
Sciences under project No. KJCX2-YW-T03 and by the National Natural
Science Foundation of China under grant Nos. 10821061, 10733010,
10725313, and by 973 Program of China under grant 2009CB824800.

\end{document}